# Effect of local stress on accurate modeling of bacterial outer membranes using all-atom molecular dynamics

Emad Pirhadi[1], Juan M. Vanegas[†,2], Mithila N. Farin[1], Jeffrey W. Schertzer[3], Xin Yong[1]*

[1]Department of Mechanical Engineering, Binghamton University, Binghamton, New York

[2]Department of Biochemistry and Biophysics, Oregon State University, Corvallis, Oregon

[3]Department of Biological Sciences, Binghamton University, Binghamton, New York

## ABSTRACT

Biological membranes are fundamental components of living organisms that play an undeniable role in their survival. Molecular dynamics (MD) serves as an essential computational tool for studying biomembranes on molecular and atomistic scales. The status quo of MD simulations of biomembranes studies a nanometer-sized membrane patch periodically extended under periodic boundary conditions (PBC). In nature, membranes are usually composed of different lipids in their two layers (referred to as leaflets). This compositional asymmetry imposes a fixed ratio of lipid numbers between the two leaflets in a periodically constrained membrane, which needs to be set appropriately. The widely adopted methods of defining leaflet lipid ratio suffer from the lack of control over the mechanical tension of each leaflet, which could significantly influence research findings. In this study, we investigate the role of membrane-building protocol and the resulting initial stress state on the interaction between small molecules and asymmetric membranes. We model the outer membrane of *Pseudomonas aeruginosa* bacteria using two different building protocols and probe their interactions with the Pseudomonas Quinolone Signal (PQS). Our results show that differential stress could shift the position of free energy minimum for the PQS molecule between the two leaflets of the asymmetric membrane. This work provides critical insights into the relationship between the initial per-leaflet tension and the spontaneous intercalation of PQS.

*Email: xyong@binghamton.edu

[†]Email: vanegasj@oregonstate.edu

# INTRODUCTION

Biological membranes are present in all domains of life, separating the cell interior from the external environment or defining different cell organelles. As selectively permeable barriers, they are involved in various cellular processes such as compartmentalization, signaling, transport and trafficking, sensing, metabolism, and overall regulation of many biological functions.[1–3] These indispensable thin films are comprised of various lipids and proteins, asymmetrically distributed between the two bilayer leaflets.[4] An important and widespread example is the outer membrane (OM) of Gram-negative bacteria, which is responsible for many of their biological traits.[5,6] While the inner leaflet of the OM is composed of common phospholipids (PLs), the abundance of lipopolysaccharides (LPS) in the outer leaflet renders the membrane highly asymmetric in leaflet lipid composition.[7] The complex structure of LPS helps shield Gram-negative bacteria against environmental threats. The structure consists of a diverse and variable length O-antigen polysaccharide in the outermost region, a more conserved oligosaccharide "core" region proximal to the membrane, and most importantly, an endotoxic Lipid A region anchored in the outer leaflet of the OM.[8] Bacterial membranes have been studied extensively in the last few decades.[9–12] Nevertheless, the undeniable role of OM in the survival of Gram-negative bacteria, including its interaction with the human immune system and contributions to bacterial pathogenesis, requires that we continuously develop and refine tools to investigate how the membrane will interact with important molecules.[13–15]

Unfortunately, state-of-the-art experimental tools still face numerous challenges in probing biomembranes and their interactions with small molecules, proteins, and nanomaterials on decreasing length and time scales.[16–18] This limitation stimulates the rapid advance in computational modeling of biomembranes, particularly molecular dynamics (MD) simulations, which offer nanoscale spatiotemporal resolutions and provide detailed information on molecular interactions not easily accessible to physical experiments.[19,20] Rapid advances in high-performance computing made it possible to use MD to investigate realistic cell membranes with diverse and complex compositions. However, the information obtained from MD requires discreet interpretation since the accuracy and reliability of these models are highly dependent on the modeling process and system parameterization. Examples of the essential parameters include the number of molecules, the representation of chemical structures (e.g., all-atom, united-atom, or coarse-grained), force fields, and electrical charge distributions.[21–23] A poorly chosen set of system parameters could significantly influence the results of MD simulations, leading to defective conclusions.[24,25] The sensitivity of MD systems could bias the understanding of physical phenomena of interest and compromise the outcomes of a research project. Therefore, it is crucial to understand the influence of critical parameters in the MD modeling of biomembranes.

Differential stress, defined as the difference between the tension of membrane leaflets, has attracted considerable attention in the community in the past few years. Hossein and Deserno reported that the differential stress could significantly influence the mechanical properties of a compositionally asymmetric membrane.[26] This behavior has also been demonstrated in recent experiments on free-standing asymmetric membranes.[27] Realistic MD models of biological membranes need to include lipid asymmetry and thus could possess pre-existing differential stress due to asymmetric lipid packing or suppressed spontaneous curvature in the periodic simulation domain.[28] Thus, the initial conditions would significantly influence the differential stress and, consequently, the mechanical properties of the model membrane. The initial stress state of the membrane due to the coupling between intrinsic bending and asymmetric lipid packing can also influence the interactions with membrane-active components such as small molecules, nanoparticles, peptides, and proteins. This emerging issue has prompted the development of new methods for rationally building an asymmetric membrane with controlled per-leaflet tension.[29,30] This work aims to elucidate the effects of building protocol as a system parameter often overlooked for accurately modeling physiological relevant OMs. Specifically, we characterize the stress states of a model OM of *Pseudomonas aeruginosa* bacteria constructed using different membrane-building protocols and systematically probe how differential stress affects the membrane interaction with signaling molecules. The findings will provide guidelines for future simulation studies of asymmetric membranes.

## MATERIALS AND METHODS

### Outer membrane and signaling molecule

Lipid A has a disaccharide backbone acylated with four to eight fatty acid tails. This potent activator of the innate immune system is also accountable for the toxic effects of Gram-negative bacteria.[31] In this model OM, the outer leaflet was composed of a hexa-acylated *P. aeruginosa*

Lipid A corresponding to the PA14 strain (see Figure 1). The lipid composition of the inner leaflet was selected based on previous experimental lipidome analysis, which reported that the PLs in *P. aeruginosa* OM mostly have one saturated and one unsaturated acyl chain.[32] The fatty acid profiles showed that the predominant pair is palmitic acid (C16:0) and oleic acid (C18:1), and their molar ratio is approximately 1:1.[33,34] From the reported PL distributions, phosphoethanolamine and phosphoglycerol (PG) are the most abundant head groups with the PE / PG ratio of 2.0 ~ 2.2.[32,34] Thus, the inner leaflet of our model OM was constructed by mixing 1-palmitoyl-2-oleoyl-sn-glycero-3-phosphoethanolamine (POPE) and 1-palmitoyl-2-oleoyl-sn-glycero-3-phosphoglycerol (POPG) at a ratio of 2.2. The bilayer was solvated in water. The system also contained corresponding numbers of calcium ($Ca^{2+}$) and sodium ($Na^+$) ions as counter ions to neutralize Lipid A and POPG, respectively. Additional NaCl at 150 mM concentration was added to mimic physiological conditions.[35] The 2-heptyl-3-hydroxy-4-quinolone (Figure 1 (d)) named *Pseudomonas* quinolone signal (PQS) is a self-produced quorum sensing molecule with multiple functionalities in the survival of *P. aeruginosa* bacteria. This study explores PQS interaction with the model OM and its quantitative effect on membrane stress.

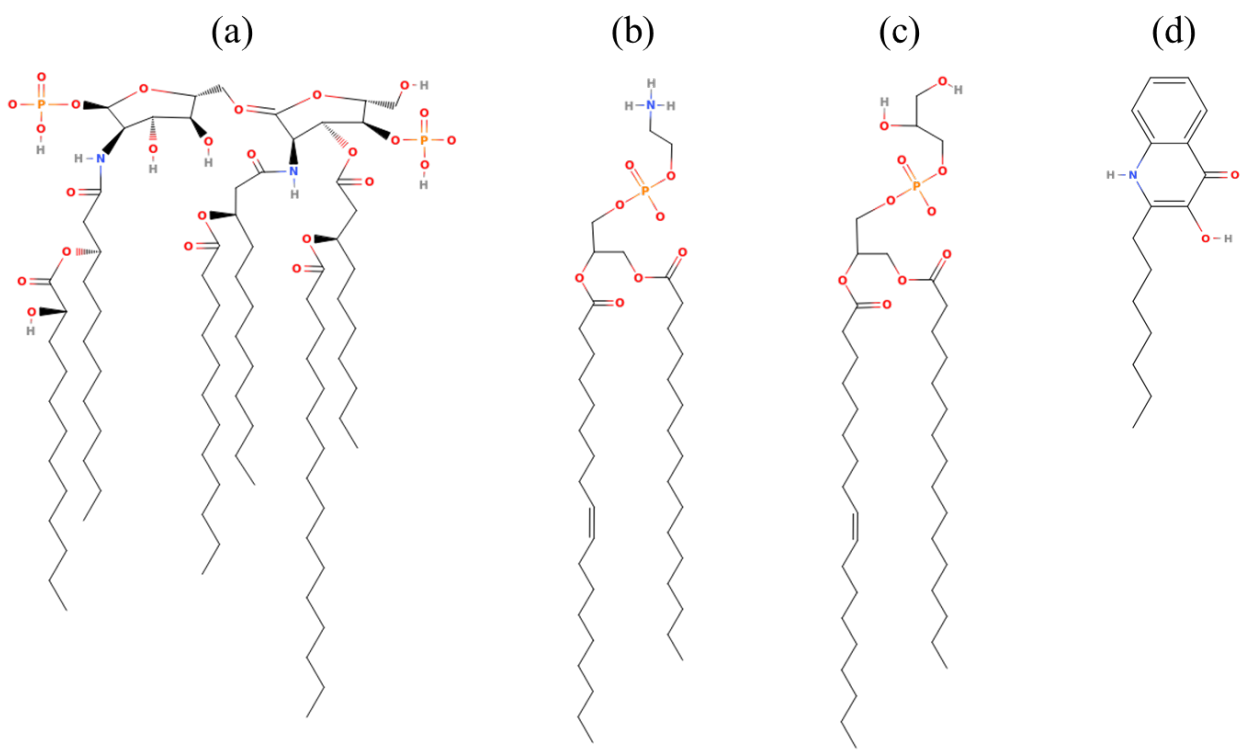

**Figure 1.** Chemical structures of (a) the *P. aeruginosa* PA14 Lipid A, (b) POPE, (c) POPG, and (d) PQS in the model system.

**Force field, Lipid A protonation state, and equilibration process**

The CHARMM36 force field[36,37] was selected to perform all-atom molecular dynamics simulation (AAMD). We modified the standard CHARMM36 parameter set using $Ca^{2+}$ NBFIX[35,38] and $Na^+$ CUFIX.[39] These nonbonded interaction Fixes improve the accuracy of ion-ion and ion-lipid interactions by changing the minimum distance of the Lennard-Jones potential defined between specific atom pairs. The net charge of Lipid A phosphate group was set to be -1, following the suggestion of a recent study by Rice *et al.*[40] To be consistent with the membrane force field, we employed the CHARMM General Force Field[41,42] to parameterize the force field and partial charge distribution of PQS.

The CHARMM-GUI membrane builder[43,44] was used to generate the initial configuration of the model membrane and PQS. All simulations in this study were carried out using the GROMACS 2021.3 package.[45,46] The membrane geometry was optimized using the steepest descent energy minimization algorithm, followed by an isothermal–isochoric (NVT) equilibration of 100 ps at 310.15 K with the velocity-rescaling thermostat.[47] The lipids and solvent were separately coupled. After that, an isothermal–isobaric (NPT) equilibration was performed for 1 ns at 1 bar with semi-isotropic pressure coupling using the Berendsen barostat while maintaining the temperature through the Berendsen thermostat.[48] The final equilibration was performed for 500-1000 ns in the NPT ensemble at 310.15 K and 1 bar with semi-isotropic pressure coupling using the Parrinello-Rahman algorithm[49] together with the Nosé-Hoover thermostat.[50] All bonds were constrained by the LINCS algorithm.[51] The cutoff of short-range van der Waals and electrostatic interactions was set to 1.2 nm with the neighbor list updated every 20 time steps. Long-range electrostatic interactions were calculated using the particle-mesh Ewald method.[52] The time step for the leap-frog integrator was set to 2 fs.

**Barostat: Parrinello-Rahman versus Stochastic Cell Rescaling**

Performing MD simulations at a constant pressure requires a numerical algorithm to control the system stress, referred to as a barostat. It has been shown that the most commonly used Parrinello-Rahman (PR) barostat could cause numerical artifacts in specific simulations.[53] The PR algorithm is based on the second Piola-Kirchhoff stress, which has a negligible difference from the true (Cauchy) stress under regular conditions with zero stress and trivial deformation. However, this difference increases as deformation increases, implying that the PR barostat may apply undesirable stress to a highly deformed system. The difference in stress measures could be problematic because the area compressibility measurement in this study (described in detail later) requires deforming the membrane from its relaxed state. We test the performance of PR barostat for a deformed membrane under tension by comparing it with the Stochastic Cell Rescaling (SCR) barostat.[54] The new algorithm was introduced recently by adding a stochastic term to the Berendsen barostat to capture correct volume fluctuations. The Berendsen barostat controls true stress with adjustments of system volume by weakly coupling it to an external bath. We performed a benchmark membrane simulation in the NPT ensemble to compare the behavior of these two barostats. The membrane deformation was imposed by keeping the normal pressure ($P_N$) constant at 1 bar while changing the lateral pressure systematically. The lateral pressure is defined as $P_L = \frac{1}{2}(P_{xx} + P_{yy})$. The bilayer mechanical tension can be obtained as $\gamma = L_z(P_N - P_L)$, where $L_z$ is the dimension of the simulation box in the membrane normal ($z$) direction. Figure 2 shows that the variation of the measured mechanical tensions with respect to area strain ($\epsilon$) is comparable between PR and SCR barostats. It indicates no notable difference between the two barostats for our system with a strain of up to 42%. However, the inset of Figure 2 shows a considerable difference between the membrane's actual pressure and the reference pressure of both barostats, making it difficult to control the exact pressure of the system. It is necessary to consider this deviation during the data analysis.

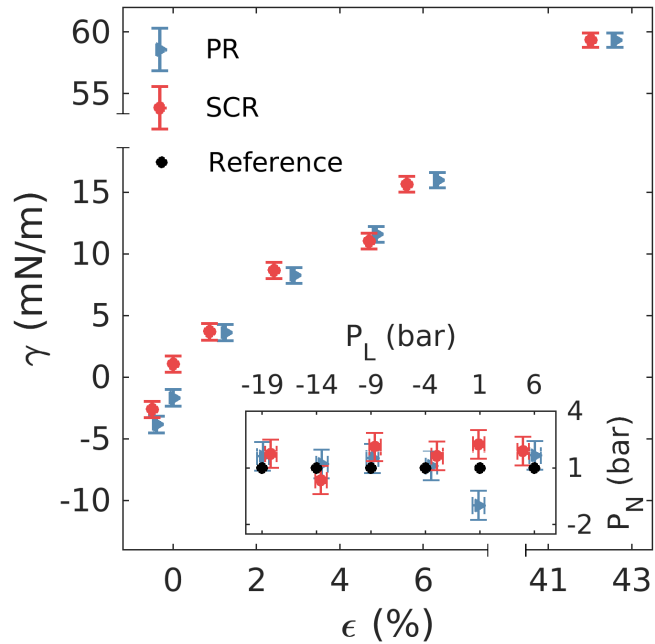

**Figure 2.** Mechanical tension as a function of membrane area strain for different barostat algorithms. Lateral pressure as a function of normal pressure is depicted in the inset.

**Local stress tensor calculation**

The GROMACS-LS,[55] a custom version of GROMACS, was employed to calculate the local stress tensor using the central force decomposition (CFD) method[56] and to obtain the lateral and normal stress profile of the bilayer. Since the current implementation of CFD in GROMACS-LS does not support reciprocal space electrostatic interactions, the long-range electrostatic contribution to the local stress has been considered only up to a cutoff radius of 3.2 nm. We examined the influences of this approximation by checking the stress profile and tension variation as the cutoff increases. Figures S1-S2 indicate that tension reaches a plateau at the Coulomb cutoff of 2.8 nm. Interestingly, tension requires a higher cutoff compared to stress profiles for convergence. For the case of stress profile convergence, it has been shown that 100 ns of analysis trajectory sampled every 5 ps is sufficient for a PL bilayer.[56] However, our preliminary investigation showed that the model OM requires significantly more data than the PL bilayer. Herein, we used 300 ns of analysis run and 600,000 frames of trajectories sampled every 5 ps for the local stress tensor and tension calculation.

**Potential of mean force: Initial configurations, umbrella sampling, and WHAM**

The potential of mean force (PMF) profile of a PQS molecule interacting with the OM was obtained using the umbrella sample method.[57] A set of initial configurations was generated by stimulating the translocation of PQS through the membrane using the steered molecular dynamics (SMD) simulation. PQS was initially positioned at 4.2 nm above the membrane in the $z$-direction. The reaction coordinate was defined as the center of mass (COM) distance between membrane and PQS in $z$. Using umbrella biasing potential with a force constant of 1000 kJ mol$^{-1}$ nm$^{-2}$, we insert and pass the PQS from the membrane with the pulling rate of 0.005 nm ps$^{-1}$.[58] In order to minimize the influence of membrane undulation on the COM distance,[59] a cylindrical section of the membrane of radius 1.5 nm located right below PQS was specified to calculate the effective membrane COM using the weighted sum of all atoms within the cylinder.[60] We selected 85 umbrella sampling windows with 0.1 nm distance between them to cover the entire range of PQS placement in the membrane. Each sampling window was equilibrated for 10 ns in the NPT ensemble. PQS can attain a multitude of orientations and conformations within the membrane, making it essential to choose an adequate time for umbrella sampling. Hence, the convergence of the PMF profile was checked by increasing the sample time by an increment of 10 ns (see Figures S5-S7). To ensure proper sampling, each window was sampled for 180-200 ns according to the membrane-building protocol. The total simulation time for the umbrella sampling process of each membrane is 15-17 μs. The weighted histogram analysis method (WHAM)[61] was employed to construct unbiased probability distribution and to obtain the PMF profile. The PQS position outside the membrane was set to be the zero reference of the PMF profile. The uncertainty in the PMF profiles was quantified using the Bayesian bootstrapping method with a 1000 resampling.[62]

## RESULTS AND DISCUSSION

**Building protocol for bilayers with compositional asymmetry**

In MD simulations, a bilayer membrane is commonly simulated in a periodically constrained system that prohibits the independent relaxation of leaflets. For membranes having asymmetric lipid compositions between leaflets, the periodic boundary conditions could lead to simulation artifacts or even an unstable model.[63] Hence, it is required to set the number of lipids in each leaflet properly to have a physical model membrane. Multiple methods have been proposed for defining the ratio of lipid numbers between the two leaflets, including matching numbers of acyl chains,[64] matching leaflet area estimated based on the ideal area per lipid of each component,[65] and matching equilibrated surface areas of leaflets (SA),[66] which is the most commonly used method. Although most of these techniques are capable of generating a stable membrane, none of them provides control over the mechanical tension ($\gamma$) within each leaflet of the obtained bilayer. In particular, when the total tension of the membrane can be maintained at zero by the barostat, individual leaflet tensions are not guaranteed to be zero. They may be equal in magnitude but with opposite signs, resulting in nonzero differential stress ($\Delta\gamma = \gamma_o - \gamma_i$) between leaflets. To address the lack of control over individual leaflet tension, Doktorova and Weinstein[29] proposed a new method capable of building an asymmetrical membrane with tensionless leaflets (0LT) (thus zero differential stress) using the relation between tension and compressibility modulus. The importance of initial conditions for asymmetric membranes also motivated the development of new periodic boundary conditions that allow interleaflet switching of specific lipid species to achieve partial chemical equilibrium and consistent mechanical properties.[30] It is important to note that the SA protocol does not regulate the sign and magnitude of differential stress. Consequently, small variations in membrane components or other related system parameters could significantly change the differential stress. In other words, if the membrane is made with the SA method, it will be almost impossible to decouple the effect of differential stress from that of a parameter of interest. Although neither necessarily relevant to biological conditions and physical experiments nor "right" in the theoretical framework, the zero-differential-stress state provides a rational reference for studying membrane mechanics in MD simulations.

**Modeling Pseudomonas aeruginosa outer membrane**

To investigate how membrane-building protocol could affect simulation results, we defined the leaflet lipid ratio of the asymmetric outer membrane using the SA method, which is the most common, and the 0LT method, which guarantees approximately zero per-leaflet tension. The equilibrated surface area of each leaflet is required for defining the ratio using the SA method. We thus modeled two symmetrical membranes, one entirely made of Lipid A and another made of POPE:POPG mixture. After proper equilibration, the areas per lipid of the symmetric membranes define the leaflet lipid ratio of the final *P. aeruginosa* outer membrane (see Table 1).

**Table 1.** Number of lipids per leaflet ($N_l$) and area per lipid (APL) from symmetric bilayer simulations. Standard deviations are given in parentheses.

| Symmetric bilayer | LPA | PL | |
|---|---|---|---|
| | | POPE | POPG |
| $N_l$ | 36 | 44 | 20 |
| APL (Å) | 158.3 (±1.6) | 59.6 (±1.4) | |

Even though the bilayer is stable, leaflets of the asymmetrical membrane made by the SA method are under tension as shown later (see Table 2). We can eliminate these tensions by adjusting the lipid ratio using the 0LT method. This building protocol assumes linear elasticity of the bilayer and defines the ideal lipid ratio ($\frac{N_o}{N_i}$) via the equation $\frac{N_o}{N_i} = \frac{N_o'(\gamma_o + K_A)}{N_i'(\gamma_i + K_A)}$. Here, $N_o$ and $N_i$ are the respective numbers of corresponding lipids in the outer and inner leaflet, which lead to zero per-leaflet tension. The calculation requires the area compressibility ($K_A$) and per-leaflet tensions ($\gamma_o$ and $\gamma_i$) of a reference membrane made with an arbitrarily chosen lipid ratio ($\frac{N_o'}{N_i'}$). We selected the membrane made by the SA method as the reference.

The area compressibility measurement was conducted based on its mechanical definition via systematical deformation of the membrane.[67] A series of membranes were simulated in the NPAT ensemble by varying the lateral dimension of the simulation box (i.e., the membrane area) while keeping the normal pressure constant at 1 bar. Figure 3 shows the variations in mechanical tension as the membrane deformation increases. The area compressibility can be obtained by fitting this curve to the equation $\left(\frac{\partial \gamma}{\partial A}\right)_T = \frac{K_A}{A_0}$. Considering the importance of $K_A$ in determining the lipid ratio of the 0LT membrane, we performed these simulations with both PR and SCR barostats, which provided us with comparable values of 260.9 mN/m and 266.9 mN/m, respectively.

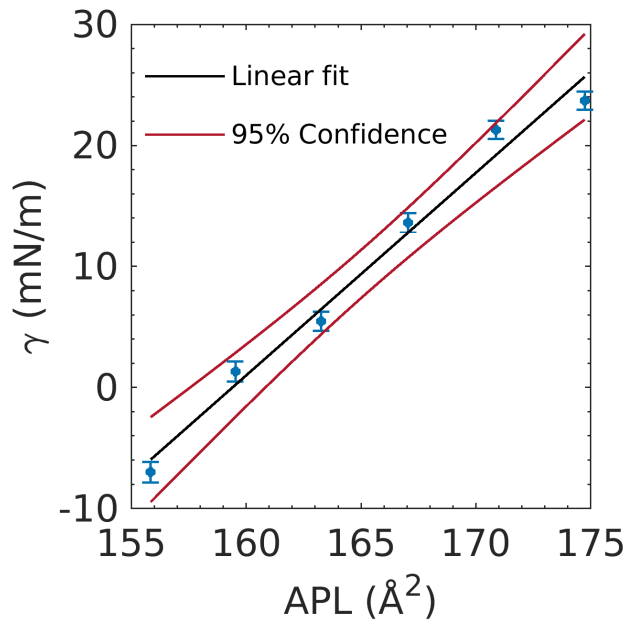

**Figure 3.** Mechanical tension as a function of area per Lipid A from the NPAT simulations.

**Calculation of per-leaflet tension.** The mechanical tension of each leaflet was obtained by $\gamma = \int_0^{\pm\infty}[P_N(z) - P_L(z)]\,dz$, where the range of integration is defined from the bilayer midplane to the bulk water. Contrary to the lateral pressure, calculating the normal pressure profile is not a trivial task in atomistic MD simulations. Many methods, including the commonly used Harasima contour, fail to deliver a constant normal pressure throughout the membrane, which is required for the mechanical equilibrium condition. To circumvent this issue, the normal pressure has been simplified by effectively averaging the lateral pressure profile using the equation $P_N = \frac{1}{L_N}\int_{-\infty}^{+\infty} P_L(z)\,dz$, which guarantees a constant normal pressure.[68] This equation has been used many times in previous studies,[29,69,70] assuming that the total mechanical tension is always zero. Our barostat evaluation indicates that this is not necessarily the case - the measured total tension of the membrane could deviate from the zero reference value. Alternatively, the CFD method included in the GROMACS-LS can obtain a constant normal pressure profile with acceptable precision (see Figure 4).

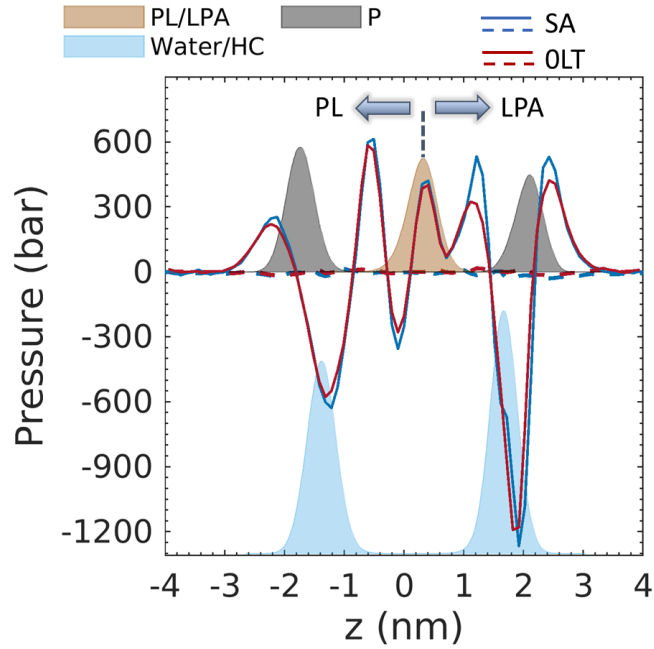

**Figure 4.** Pressure profiles as functions of the distance from the membrane's COM along the membrane normal direction. Lateral and normal pressure profiles are represented by solid and dashed lines, respectively. Scaled density profiles are depicted using shaded areas. The interface of water and hydrocarbon tails (Water/HC) is identified by $\rho_{water}(z) \cdot \rho_{tails}(z)$, and similarly, the Lipid A and phospholipids interface (PL/LPA) interface is presented by $\rho_{LPA\text{-}tails}(z) \cdot \rho_{PL\text{-}tails}(z)$.

To define the lipid ratio, we need a normal pressure profile, and there are two ways of calculating it as mentioned above. To ensure a proper choice, we built two distinct membranes following the 0LT protocol, each using a different way of measuring the normal pressure. In particular, we employed the CFD and GROMACS-LS for the first membrane and the frequently used $P_N = \frac{1}{L_N} \int P_L(z)\, dz$ equation for the second membrane. Table S1 shows that the variation in the differential stress between the SA and 0LT membranes is greater when the normal pressure is calculated by GROMACS-LS. The comparison indicates that implementing the 0LT method with GROMACS-LS more efficiently reduces the initial differential stress of asymmetric membranes. In addition, by using GROMACS-LS we include information about the membrane actual state in the normal pressure calculation instead of assuming an ideal system. Thus, we employ this method to calculate the normal pressure profile for the rest of this study.

Figure 4 shows that membrane-building protocol significantly influences the pressure profile. Notably, we observe a reduction in repulsive stress in the outer leaflet. In addition, Table 2 demonstrates that the membrane made using the 0LT method has lower differential stress and per-leaflet tension compared to the SA membrane as expected. It is impossible to build a finite membrane with exactly zero-tension leaflets because the desired lipid ratio can only be approximated given an integer number of lipids. Furthermore, the pressure profile calculated using GROMACS-LS is not exact due to the truncation of long-range electrostatic forces.

**Table 2.** Differential stress, per-leaflet tension, and area per lipid of the membranes made with the SA and 0LT methods. The standard errors of $\gamma$ and the standard deviation of APL are given in parentheses. The uncertainty in $\gamma$ was estimated by block averaging.

| Membrane | $\Delta\gamma$ (mN/m) | $\gamma$ (mN/m) | | APL (Å) | |
|---|---|---|---|---|---|
| | | Outer | Inner | LPA | PL |
| SA | -6.14 | -5.40 (±0.26) | 0.74 (±0.10) | 157.1 (±2.0) | 59.1 (±0.8) |
| 0LT | 1.36 | 1.31 (±0.09) | -0.05 (±0.10) | 160.0 (±1.9) | 58.9 (±0.7) |

Perception of membrane initial stress state and its impact on simulation results could be challenging by only comparing numbers and discussing properties. In the following section, we provide physical insights into this matter by exploring the interaction of an important signaling molecule with the two *P. aeruginosa* membranes made with the SA and 0LT methods. Analyses and comparisons of these systems generate a deeper understanding of how leaflets' pre-existing tension influences the behavior of a physiologically relevant system. More importantly, the results highlight the importance of exercising prudence in using the SA protocol for building a membrane.

**Outer membrane interaction with quorum sensing molecule**

Bacterial cell-cell communication ("quorum sensing") is critical for regulating population responses to the environment. PQS is a signaling molecule produced by *P. aeruginosa* that plays an essential role in bacterial communication in response to population change, including the activation of many virulence-associated programs.[71–73] This molecule has also been recognized to induce outer membrane vesicle (OMV) production through specific interactions with the *P. aeruginosa* OM.[74–76] In particular, PQS intercalation into the OM induces membrane curvature formation, which is an essential early step in OMV production.[58,77] Considering the influences of OMVs in the competitiveness and pathogenicity of many gram-negative bacterial species,[78–81] PQS has the potential as a drug target for combating various infectious diseases.

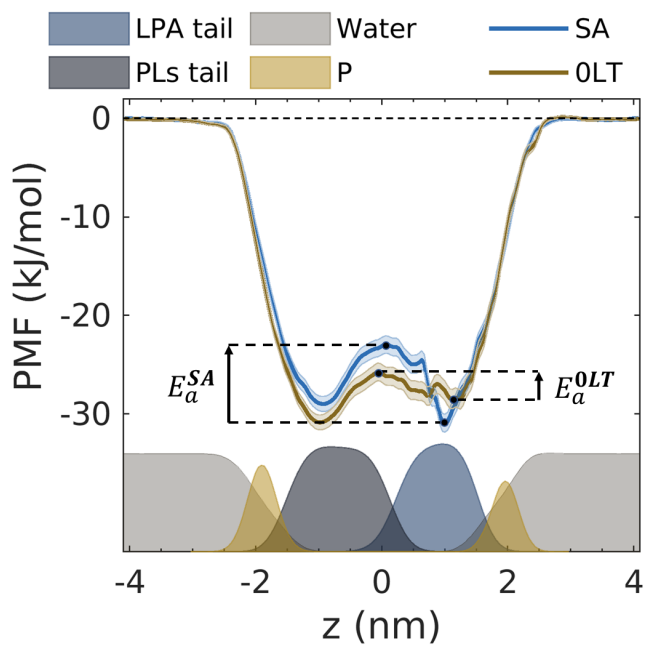

**Figure 5.** Transmembrane potential of mean force profiles along the membrane normal direction for PQS interacting with the SA and 0LT membranes. Scaled density profiles are depicted using shaded areas. $E_a$ represents the activation energy for the flip-flop of PQS from the outer to the inner leaflet. The membrane center of mass is located at $z = 0$. The shaded error region represents the standard deviation.

We characterize the PQS interaction with the model OMs through the PMF profile, which demonstrates the free energy variations of PQS at different locations relevant to the membrane. As shown in Figures S5-S7, the convergence of the PMF profile for the SA membrane requires 120 ns of data sampling for each window, while the PMF profile of the 0LT membrane converges after 80 ns. For the current system, this difference tallies up to 3.4 μs in the total simulation time for the entire umbrella sampling of PQS interaction, which reduces the computational cost substantially. Figure 5 shows that the PMF profile is negative in the membrane region independent of the building protocol, which denotes that PQS would spontaneously insert into the membrane. For both membranes, there are two local minima located in the tail region of leaflets near the position where water density completely vanishes, suggesting the insertion is attributed to the strong hydrophobic nature of the alkyl chain. The PMF profile of the 0LT membrane exhibits a deeper well in the inner leaflet corresponding to a global free energy minimum. Interestingly, it is the outer leaflet that has a deeper well in the SA membrane, suggesting that an unbalanced tension influences the preferential position of a small molecule within the membrane. In addition, the comparison between the SA and 0LT membrane highlights the energy barrier separating the two local minima is also affected by the membrane stress state. More specifically, the activation energy for the outer-to-inner leaflet movement is larger for the membrane made with the SA method, indicating that nonzero differential stress hinders PQS unrestrained motion within the membrane. This energy barrier would influence the probability and frequency of PQS transmembrane diffusion (flip-flop) between the leaflets and therefore modulates the membrane mechanical properties.[82] Figure 6 shows that the average total number of hydrogen bonds ($N_{Hb}$) between PQS and the membrane is significantly influenced by the stress state and the membrane made with the SA method has a larger $N_{Hb}$. This explains the difference in the magnitude of PMF energy barrier between the two membranes. These results provide critical insights into how having an asymmetric membrane with an incorrectly balanced tension could lead to incorrectly estimating the translocation free energy of a molecule across the bilayer. Furthermore, the *z*-positions of PMF local minima do not coincide with peak positions of $N_{Hb}$. Instead, these favorable positions of PQS are shifted toward the center to avoid any interaction with water.

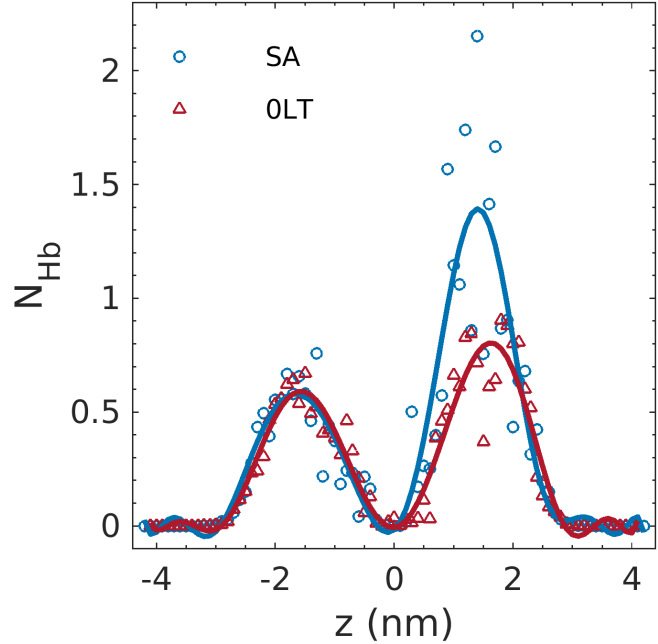

**Figure 6.** The total average number of hydrogen bonds between the PQS and membranes as a function of PQS distance from the COM of the membrane in the normal coordinate. The solid lines represent the spline interpolation fit of the data as guides for the eye.

Our current model membrane possesses a differential stress of 6.14 mN/m if we use the SA method. Compared with the 0LT method, this differential stress induces a variation of approximately 5 kJ/mol (~2 $k_\text{B}T$) in the outer-to-inner flip-flop activation energy ($E_a$) and a 0.6 change in the number of hydrogen bonds. Here we should emphasize again that the SA method does not control the sign or the magnitude of pre-existing differential stress. This justifies using the 0LT building protocol instead of the SA method in similar studies to obtain a more consistent reference system.

**Table 3.** Differential stress, per-leaflet tension, and area per lipid of the membranes made with the SA and 0LT method before and after insertion of PQS to the outer leaflet. The standard errors of $\gamma$ and the standard deviation of APL are given in parentheses. The uncertainty in $\gamma$ was estimated by block averaging.

| Membrane | $\Delta\gamma$ | $\gamma\ (mN/m)$ | APL (Å) |
| --- | --- | --- | --- |

|         | $(mN/m)$ | Outer | Inner | LPA | PL |
|---------|----------|-------|-------|-----|-----|
| *SA*    | *-6.14*  | *-5.40* | *0.74* | *157.1* | *59.1* |
| **SA+PQS** | **-2.05** | **-1.04** (±0.08) | **1.01** (±0.16) | **157.7** (±2.0) | **59.4** (±0.7) |
| *0LT*   | *1.36*   | *1.31* | *-0.05* | *160.0* | *58.9* |
| **0LT+PQS** | **4.48** | **0.20** (±0.12) | **-4.28** (±0.16) | **158.3** (±2.0) | **58.2** (±0.7) |

**Membrane with a pre-inserted PQS**

In biological systems, PQS molecules are generally believed to be introduced to the membrane from the outer leaflet side.[77,83] To further investigate this phenomenon, we insert a PQS to the outer leaflet of both the SA and 0LT membranes at the locations of free energy minima obtained from the PMF profiles. The relative position is maintained by restraining the COM distance of PQS and membrane in the *z* direction using an umbrella potential. The stress profile of the system as well as per-leaflet tensions are calculated using the method described before. Figure 7 indicates that PQS insertion changes the lateral pressure of the 0LT membrane noticeably and simultaneously reduces the APL slightly, rendering the system more crowded (see Table 3). This deviation from the equilibrium state of the system induces considerable differential stress within the membrane. In the previous section, we showed that higher differential stress could reduce the probability of PQS flip-flop. This could lead to the accumulation of PQS molecules in the outer leaflet of *P. aeruginosa* bacteria OM. The membrane may try to release this excessive tension by forming curvature, which would initiate the process of OMV production.[77] Although PQS is located within the outer leaflet, the inner leaflet tension reduces more significantly, showing a more pronounced response to the insertion. We believe the large number of hydrogen bonds formed between PQS and the outer leaflet will compensate for the effect of increased molecular packing on the tension. Moreover, the addition of PQS will disturb the balance of the membrane torque (second moment of the pressure profile) under the constraint of periodic boundary conditions, which could result in inconsistent variations in tension between leaflets. We encourage future studies to investigate this behavior by measuring the membrane torque, which is beyond the scope of the current work.

Interestingly, PQS insertion has a minor influence on the lateral pressure profile of the SA membrane (see Figure S8) and in contrast to the 0LT membrane, reduces the differential stress. Since leaflets of the SA membrane are initially under tension, PQS addition could either exacerbate that or help balance the tension. Insertion of PQS or any other external molecules into a membrane made with the SA method would stimulate an unpredictable response due to the fact that this method generates a membrane with an unknown initial stress state.

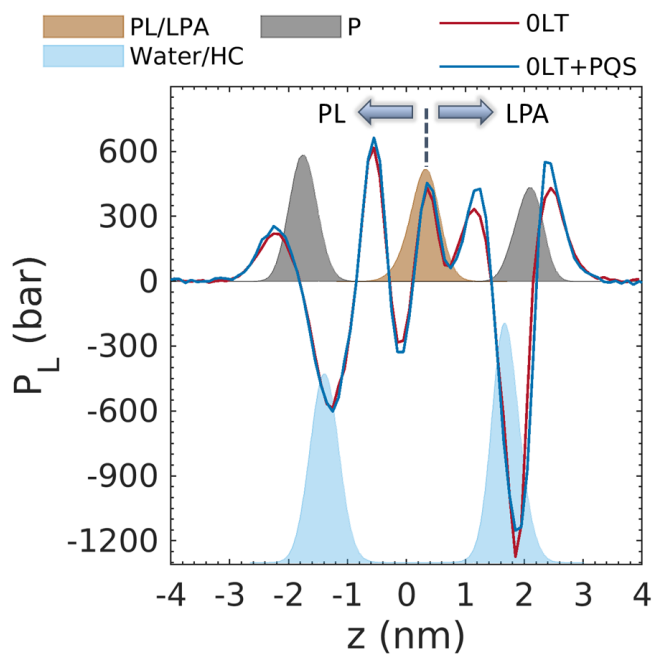

**Figure 7.** Lateral pressure profiles of the 0LT membrane before and after the insertion of PQS as functions of the distance from the membrane's COM along the membrane normal direction. Scaled density profiles are depicted using shaded areas. The interface of water and hydrocarbon tails (Water/HC) is identified by $\rho_{water}(z) \cdot \rho_{tails}(z)$, and similarly, the Lipid A and phospholipids interface (PL/LPA) interface is presented by $\rho_{LPA\text{-}tails}(z) \cdot \rho_{PL\text{-}tails}(z)$.

## CONCLUSION

The combination of asymmetry in membrane lipid composition with the periodic boundary condition necessitates the specification of a fixed ratio of lipids between leaflets. Most traditional protocols defining this ratio, including the commonly used surface-area-matching method, create membranes with pre-existing per-leaflet tension. Even though physical per-leaflet tensions of biomembranes are still unknown, the unawareness of the numerical approach that generates this tension is alarming. In this study, we extended previous studies in probing this newly recognized issue and demonstrated the significance of the initial stress state on the membrane interactions with signaling molecules. Our data shows that a lack of control over the per-leaflet tensions of a membrane could cause significant inaccuracy in the measurement of stress profiles, leaflets' tension, number of hydrogen bonds, and magnitude of potential of mean force. In addition, variation in differential stress showed a notable influence on the membrane response to the insertion of small molecules. Essentially, with an unpredictable per-leaflet tension, it would be difficult to recognize the pure dependency of the generated data on the parameter of the study. We suggest using methods that allow us to control leaflet tension instead of the surface-area-matching or other previously used methods to at least regulate the influence of the initial stress state. We also discuss the limitation of current numerical tools in measuring the local stress tensors. We encourage future studies to proceed with these insights to improve the realistic modeling of biological membranes.

## ACKNOWLEDGMENTS

X. Y. and J. W. S. acknowledge the support from the National Institute of General Medical Sciences through grant R15GM135862. Computing time was provided by the Extreme Science and Engineering Discovery Environment (XSEDE) through allocations MCB190029 and BIO220055, which is supported by the National Science Foundation grant number ACI-1548562. This work also used the Center for Functional Nanomaterials, which is a U.S. DOE Office of Science Facility, at Brookhaven National Laboratory under Contract No. DESC0012704. J. M. V. acknowledges the support of the National Science Foundation through Grant No. CHE-1944892.

**Supporting Information:**

The supporting information is available free of charge at:

Supplementary Figures and Table: minimum Coulomb cutoff required by GROMACS-LS for accurate stress profile calculation; adequate sampling time required for the convergence of PMF profiles; comparison of two methods of normal pressure calculation

# REFERENCES

(1) Musille, P. M.; Kohn, J. A.; Ortlund, E. A. Phospholipid - Driven Gene Regulation. *FEBS Lett.* **2013**, *587*, 1238–1246.

(2) Botelho, A. V.; Huber, T.; Sakmar, T. P.; Brown, M. F. Curvature and Hydrophobic Forces Drive Oligomerization and Modulate Activity of Rhodopsin in Membranes. *Biophys. J.* **2006**, *91*, 4464–4477.

(3) Yang, Y.; Lee, M.; Fairn, G. D. Phospholipid Subcellular Localization and Dynamics. *J. Biol. Chem.* **2018**, *293*, 6230–6240.

(4) van Meer, G.; Voelker, D. R.; Feigenson, G. W. Membrane Lipids: Where They Are and How They Behave. *Nat. Rev. Mol. Cell Biol.* **2008**, *9*, 112–124.

(5) Silhavy, T. J.; Kahne, D.; Walker, S. The Bacterial Cell Envelope. *Cold Spring Harb. Perspect. Biol.* **2010**, *2*, a000414–a000414.

(6) Rojas, E. R.; Billings, G.; Odermatt, P. D.; Auer, G. K.; Zhu, L.; Miguel, A.; Chang, F.; Weibel, D. B.; Theriot, J. A.; Huang, K. C. The Outer Membrane Is an Essential Load-Bearing Element in Gram-Negative Bacteria. *Nature* **2018**, *559*, 617–621.

(7) Nikaido, H. Molecular Basis of Bacterial Outer Membrane Permeability Revisited. *Microbiol. Mol. Biol. Rev.* **2003**, *67*, 593–656.

(8) Bos, M. P.; Robert, V.; Tommassen, J. Biogenesis of the Gram-Negative Bacterial Outer Membrane. *Annu. Rev. Microbiol.* **2007**, *61*, 191–214.

(9) Koebnik, R.; Locher, K. P.; Van Gelder, P. Structure and Function of Bacterial Outer Membrane Proteins: Barrels in a Nutshell. *Mol. Microbiol.* **2000**, *37*, 239–253.

(10) Schirmer, T. General and Specific Porins from Bacterial Outer Membranes. *J. Struct. Biol.* **1998**, *121*, 101–109.

(11) Kuehn, M. J.; Kesty, N. C. Bacterial Outer Membrane Vesicles and the Host–Pathogen Interaction. *Genes Dev.* **2005**, *19*, 2645–2655.

(12) HANCOCK, R. The Bacterial Outer Membrane as a Drug Barrier. *Trends Microbiol.*

**1997**, *5*, 37–42.

(13) Tenover, F. C. Mechanisms of Antimicrobial Resistance in Bacteria. *Am. J. Med.* **2006**, *119*, S3–S10.

(14) Weiner-Lastinger, L. M.; Abner, S.; Edwards, J. R.; Kallen, A. J.; Karlsson, M.; Magill, S. S.; Pollock, D.; See, I.; Soe, M. M.; Walters, M. S.; Dudeck, M. A. Antimicrobial-Resistant Pathogens Associated with Adult Healthcare-Associated Infections: Summary of Data Reported to the National Healthcare Safety Network, 2015–2017. *Infect. Control Hosp. Epidemiol.* **2020**, *41*, 1–18.

(15) Rice, L. B. Federal Funding for the Study of Antimicrobial Resistance in Nosocomial Pathogens: No ESKAPE. *J. Infect. Dis.* **2008**, *197*, 1079–1081.

(16) Lyman, E.; Hsieh, C.-L.; Eggeling, C. From Dynamics to Membrane Organization: Experimental Breakthroughs Occasion a "Modeling Manifesto." *Biophys. J.* **2018**, *115*, 595–604.

(17) Nickels, J. D.; Smith, J. C.; Cheng, X. Lateral Organization, Bilayer Asymmetry, and Inter-Leaflet Coupling of Biological Membranes. *Chem. Phys. Lipids* **2015**, *192*, 87–99.

(18) Deleu, M.; Crowet, J.-M.; Nasir, M. N.; Lins, L. Complementary Biophysical Tools to Investigate Lipid Specificity in the Interaction between Bioactive Molecules and the Plasma Membrane: A Review. *Biochim. Biophys. Acta - Biomembr.* **2014**, *1838*, 3171–3190.

(19) McCammon, J. A.; Gelin, B. R.; Karplus, M. Dynamics of Folded Proteins. *Nature* **1977**, *267*, 585–590.

(20) Ingólfsson, H. I.; Arnarez, C.; Periole, X.; Marrink, S. J. Computational 'Microscopy' of Cellular Membranes. *J. Cell Sci.* **2016**.

(21) Mackerell, A. D. Empirical Force Fields for Biological Macromolecules: Overview and Issues. *J. Comput. Chem.* **2004**, *25*, 1584–1604.

(22) Riniker, S. Fixed-Charge Atomistic Force Fields for Molecular Dynamics Simulations in the Condensed Phase: An Overview. *J. Chem. Inf. Model.* **2018**, *58*, 565–578.

(23) Marrink, S. J.; Corradi, V.; Souza, P. C. T.; Ingólfsson, H. I.; Tieleman, D. P.; Sansom, M. S. P. Computational Modeling of Realistic Cell Membranes. *Chem. Rev.* **2019**, *119*, 6184–6226.

(24) Patra, M.; Karttunen, M.; Hyvönen, M. T.; Falck, E.; Lindqvist, P.; Vattulainen, I. Molecular Dynamics Simulations of Lipid Bilayers: Major Artifacts Due to Truncating Electrostatic Interactions. *Biophys. J.* **2003**, *84*, 3636–3645.

(25) Feller, S. E.; Pastor, R. W. On Simulating Lipid Bilayers with an Applied Surface Tension: Periodic Boundary Conditions and Undulations. *Biophys. J.* **1996**, *71*, 1350–1355.

(26) Hossein, A.; Deserno, M. Spontaneous Curvature, Differential Stress, and Bending Modulus of Asymmetric Lipid Membranes. *Biophys. J.* **2020**, *118*, 624–642.

(27) Liu, P.; Zabala-Ferrera, O.; Beltramo, P. J. Fabrication and Electromechanical Characterization of Freestanding Asymmetric Membranes. *Biophys. J.* **2021**, *120*, 1755–1764.

(28) Różycki, B.; Lipowsky, R. Spontaneous Curvature of Bilayer Membranes from Molecular Simulations: Asymmetric Lipid Densities and Asymmetric Adsorption. *J. Chem. Phys.* **2015**, *142*, 054101.

(29) Doktorova, M.; Weinstein, H. Accurate In Silico Modeling of Asymmetric Bilayers Based on Biophysical Principles. *Biophys. J.* **2018**, *115*, 1638–1643.

(30) Park, S.; Im, W.; Pastor, R. W. Developing Initial Conditions for Simulations of Asymmetric Membranes: A Practical Recommendation. *Biophys. J.* **2021**, *120*, 5041–5059.

(31) Takayama, K.; Qureshi, N.; Ribi, E.; Cantrell, J. L. Separation and Characterization of Toxic and Nontoxic Forms of Lipid A. *Clin. Infect. Dis.* **1984**, *6*, 439–443.

(32) Benamara, H.; Rihouey, C.; Abbes, I.; Ben Mlouka, M. A.; Hardouin, J.; Jouenne, T.; Alexandre, S. Characterization of Membrane Lipidome Changes in Pseudomonas Aeruginosa during Biofilm Growth on Glass Wool. *PLoS One* **2014**, *9*, e108478.

(33) Chao, J.; Wolfaardt, G. M.; Arts, M. T. Characterization of Pseudomonas Aeruginosa Fatty Acid Profiles in Biofilms and Batch Planktonic Cultures. *Can. J. Microbiol.* **2010**, *56*, 1028–1039.

(34) TASHIRO, Y.; INAGAKI, A.; SHIMIZU, M.; ICHIKAWA, S.; TAKAYA, N.; NAKAJIMA-KAMBE, T.; UCHIYAMA, H.; NOMURA, N. Characterization of Phospholipids in Membrane Vesicles Derived from Pseudomonas Aeruginosa. *Biosci. Biotechnol. Biochem.* **2011**, *75*, 605–607.

(35) Kim, S.; Patel, D. S.; Park, S.; Slusky, J.; Klauda, J. B.; Widmalm, G.; Im, W. Bilayer Properties of Lipid A from Various Gram-Negative Bacteria. *Biophys. J.* **2016**, *111*, 1750–1760.

(36) Klauda, J. B.; Venable, R. M.; Freites, J. A.; O'Connor, J. W.; Tobias, D. J.; Mondragon-Ramirez, C.; Vorobyov, I.; MacKerell, A. D.; Pastor, R. W. Update of the CHARMM All-Atom Additive Force Field for Lipids: Validation on Six Lipid Types. *J. Phys. Chem. B* **2010**, *114*, 7830–7843.

(37) Vanommeslaeghe, K.; MacKerell, A. D. CHARMM Additive and Polarizable Force Fields for Biophysics and Computer-Aided Drug Design. *Biochim. Biophys. Acta - Gen. Subj.* **2015**, *1850*, 861–871.

(38) Han, K.; Venable, R. M.; Bryant, A.-M.; Legacy, C. J.; Shen, R.; Li, H.; Roux, B.; Gericke, A.; Pastor, R. W. Graph–Theoretic Analysis of Monomethyl Phosphate Clustering in Ionic Solutions. *J. Phys. Chem. B* **2018**, *122*, 1484–1494.

(39) Yoo, J.; Aksimentiev, A. Improved Parametrization of Li + , Na + , K + , and Mg 2+ Ions for All-Atom Molecular Dynamics Simulations of Nucleic Acid Systems. *J. Phys. Chem. Lett.* **2012**, *3*, 45–50.

(40) Rice, A.; Rooney, M. T.; Greenwood, A. I.; Cotten, M. L.; Wereszczynski, J. Lipopolysaccharide Simulations Are Sensitive to Phosphate Charge and Ion Parameterization. *J. Chem. Theory Comput.* **2020**, *16*, 1806–1815.

(41) Vanommeslaeghe, K.; Hatcher, E.; Acharya, C.; Kundu, S.; Zhong, S.; Shim, J.; Darian, E.; Guvench, O.; Lopes, P.; Vorobyov, I.; Mackerell, A. D. CHARMM General Force

Field: A Force Field for Drug-like Molecules Compatible with the CHARMM All-Atom Additive Biological Force Fields. *J. Comput. Chem.* **2009**, NA-NA.

(42) Vanommeslaeghe, K.; Raman, E. P.; MacKerell, A. D. Automation of the CHARMM General Force Field (CGenFF) II: Assignment of Bonded Parameters and Partial Atomic Charges. *J. Chem. Inf. Model.* **2012**, *52*, 3155–3168.

(43) Jo, S.; Kim, T.; Iyer, V. G.; Im, W. CHARMM-GUI: A Web-Based Graphical User Interface for CHARMM. *J. Comput. Chem.* **2008**, *29*, 1859–1865.

(44) Jo, S.; Cheng, X.; Lee, J.; Kim, S.; Park, S.; Patel, D. S.; Beaven, A. H.; Lee, K. Il; Rui, H.; Park, S.; Lee, H. S.; Roux, B.; MacKerell, A. D.; Klauda, J. B.; Qi, Y.; Im, W. CHARMM-GUI 10 Years for Biomolecular Modeling and Simulation. *J. Comput. Chem.* **2017**, *38*, 1114–1124.

(45) Berendsen, H. J. C.; van der Spoel, D.; van Drunen, R. GROMACS: A Message-Passing Parallel Molecular Dynamics Implementation. *Comput. Phys. Commun.* **1995**, *91*, 43–56.

(46) Abraham, M. J.; Murtola, T.; Schulz, R.; Páll, S.; Smith, J. C.; Hess, B.; Lindahl, E. GROMACS: High Performance Molecular Simulations through Multi-Level Parallelism from Laptops to Supercomputers. *SoftwareX* **2015**, *1–2*, 19–25.

(47) Woodcock, L. V. Isothermal Molecular Dynamics Calculations for Liquid Salts. *Chem. Phys. Lett.* **1971**, *10*, 257–261.

(48) Berendsen, H. J. C.; Postma, J. P. M.; van Gunsteren, W. F.; DiNola, A.; Haak, J. R. Molecular Dynamics with Coupling to an External Bath. *J. Chem. Phys.* **1984**, *81*, 3684–3690.

(49) Parrinello, M.; Rahman, A. Polymorphic Transitions in Single Crystals: A New Molecular Dynamics Method. *J. Appl. Phys.* **1981**, *52*, 7182–7190.

(50) Evans, D. J.; Holian, B. L. The Nose–Hoover Thermostat. *J. Chem. Phys.* **1985**, *83*, 4069–4074.

(51) Hess, B. P-LINCS: A Parallel Linear Constraint Solver for Molecular Simulation. *J. Chem. Theory Comput.* **2008**, *4*, 116–122.

(52) Darden, T.; York, D.; Pedersen, L. Particle Mesh Ewald: An N·log(N) Method for Ewald Sums in Large Systems. *J. Chem. Phys.* **1993**, *98*, 10089–10092.

(53) Miller, R. E.; Tadmor, E. B.; Gibson, J. S.; Bernstein, N.; Pavia, F. Molecular Dynamics at Constant Cauchy Stress. *J. Chem. Phys.* **2016**, *144*, 184107.

(54) Bernetti, M.; Bussi, G. Pressure Control Using Stochastic Cell Rescaling. *J. Chem. Phys.* **2020**, *153*, 114107.

(55) Ollila, O. H. S.; Risselada, H. J.; Louhivuori, M.; Lindahl, E.; Vattulainen, I.; Marrink, S. J. 3D Pressure Field in Lipid Membranes and Membrane-Protein Complexes. *Phys. Rev. Lett.* **2009**, *102*, 078101.

(56) Vanegas, J. M.; Torres-Sánchez, A.; Arroyo, M. Importance of Force Decomposition for Local Stress Calculations in Biomembrane Molecular Simulations. *J. Chem. Theory Comput.* **2014**, *10*, 691–702.

(57) Torrie, G. M.; Valleau, J. P. Nonphysical Sampling Distributions in Monte Carlo Free-Energy Estimation: Umbrella Sampling. *J. Comput. Phys.* **1977**, *23*, 187–199.

(58) Li, A.; Schertzer, J. W.; Yong, X. Molecular Conformation Affects the Interaction of the Pseudomonas Quinolone Signal with the Bacterial Outer Membrane. *J. Biol. Chem.* **2019**, *294*, 1089–1094.

(59) Neale, C.; Pomès, R. Sampling Errors in Free Energy Simulations of Small Molecules in Lipid Bilayers. *Biochim. Biophys. Acta - Biomembr.* **2016**, *1858*, 2539–2548.

(60) Nitschke, N.; Atkovska, K.; Hub, J. S. Accelerating Potential of Mean Force Calculations for Lipid Membrane Permeation: System Size, Reaction Coordinate, Solute-Solute Distance, and Cutoffs. *J. Chem. Phys.* **2016**, *145*, 125101.

(61) Kumar, S.; Rosenberg, J. M.; Bouzida, D.; Swendsen, R. H.; Kollman, P. A. THE Weighted Histogram Analysis Method for Free-Energy Calculations on Biomolecules. I. The Method. *J. Comput. Chem.* **1992**, *13*, 1011–1021.

(62) Hub, J. S.; de Groot, B. L.; van der Spoel, D. G_wham—A Free Weighted Histogram Analysis Implementation Including Robust Error and Autocorrelation Estimates. *J. Chem.*

*Theory Comput.* **2010**, *6*, 3713–3720.

(63) Li, A.; Schertzer, J. W.; Yong, X. Molecular Dynamics Modeling of Pseudomonas Aeruginosa Outer Membranes. *Phys. Chem. Chem. Phys.* **2018**, *20*, 23635–23648.

(64) Kirschner, K. N.; Lins, R. D.; Maass, A.; Soares, T. A. A Glycam-Based Force Field for Simulations of Lipopolysaccharide Membranes: Parametrization and Validation. *J. Chem. Theory Comput.* **2012**, *8*, 4719–4731.

(65) Jo, S.; Kim, T.; Im, W. Automated Builder and Database of Protein/Membrane Complexes for Molecular Dynamics Simulations. *PLoS One* **2007**, *2*, e880.

(66) Vácha, R.; Berkowitz, M. L.; Jungwirth, P. Molecular Model of a Cell Plasma Membrane With an Asymmetric Multicomponent Composition: Water Permeation and Ion Effects. *Biophys. J.* **2009**, *96*, 4493–4501.

(67) Feller, S. E.; Pastor, R. W. Constant Surface Tension Simulations of Lipid Bilayers: The Sensitivity of Surface Areas and Compressibilities. *J. Chem. Phys.* **1999**, *111*, 1281–1287.

(68) Sodt, A. J.; Pastor, R. W. Bending Free Energy from Simulation: Correspondence of Planar and Inverse Hexagonal Lipid Phases. *Biophys. J.* **2013**, *104*, 2202–2211.

(69) Park, S.; Beaven, A. H.; Klauda, J. B.; Im, W. How Tolerant Are Membrane Simulations with Mismatch in Area per Lipid between Leaflets? *J. Chem. Theory Comput.* **2015**, *11*, 3466–3477.

(70) Khelashvili, G.; Plante, A.; Doktorova, M.; Weinstein, H. $Ca^{2+}$-Dependent Mechanism of Membrane Insertion and Destabilization by the SARS-CoV-2 Fusion Peptide. *Biophys. J.* **2021**, *120*, 1105–1119.

(71) Pesci, E. C.; Milbank, J. B. J.; Pearson, J. P.; McKnight, S.; Kende, A. S.; Greenberg, E. P.; Iglewski, B. H. Quinolone Signaling in the Cell-to-Cell Communication System of Pseudomonas Aeruginosa. *Proc. Natl. Acad. Sci.* **1999**, *96*, 11229–11234.

(72) Rutherford, S. T.; Bassler, B. L. Bacterial Quorum Sensing: Its Role in Virulence and Possibilities for Its Control. *Cold Spring Harb. Perspect. Med.* **2012**, *2*, a012427–a012427.

(73) Lee, J.; Zhang, L. The Hierarchy Quorum Sensing Network in Pseudomonas Aeruginosa. *Protein Cell* **2015**, *6*, 26–41.

(74) Mashburn, L. M.; Whiteley, M. Membrane Vesicles Traffic Signals and Facilitate Group Activities in a Prokaryote. *Nature* **2005**, *437*, 422–425.

(75) Mashburn-Warren, L.; Howe, J.; Garidel, P.; Richter, W.; Steiniger, F.; Roessle, M.; Brandenburg, K.; Whiteley, M. Interaction of Quorum Signals with Outer Membrane Lipids: Insights into Prokaryotic Membrane Vesicle Formation. *Mol. Microbiol.* **2008**, *69*, 491–502.

(76) Horspool, A. M.; Schertzer, J. W. Reciprocal Cross-Species Induction of Outer Membrane Vesicle Biogenesis via Secreted Factors. *Sci. Rep.* **2018**, *8*, 9873.

(77) Schertzer, J. W.; Whiteley, M. A Bilayer-Couple Model of Bacterial Outer Membrane Vesicle Biogenesis. *MBio* **2012**, *3*.

(78) Schwechheimer, C.; Kuehn, M. J. Outer-Membrane Vesicles from Gram-Negative Bacteria: Biogenesis and Functions. *Nat. Rev. Microbiol.* **2015**, *13*, 605–619.

(79) Lin, L.; Wang, Y.; Srinivasan, R.; Zhang, L.; Song, H.; Song, Q.; Wang, G.; Lin, X. Quantitative Proteomics Reveals That the Protein Components of Outer Membrane Vesicles (OMVs) in Aeromonas Hydrophila Play Protective Roles in Antibiotic Resistance. *J. Proteome Res.* **2022**, *21*, 1707–1717.

(80) Schertzer, J. W.; Whiteley, M. Bacterial Outer Membrane Vesicles in Trafficking, Communication and the Host-Pathogen Interaction. *Microb. Physiol.* **2013**, *23*, 118–130.

(81) Vella, B. D.; Schertzer, J. W. Understanding and Exploiting Bacterial Outer Membrane Vesicles. In *Pseudomonas*; Springer Netherlands: Dordrecht, **2015**; pp. 217–250.

(82) Bruckner, R. J.; Mansy, S. S.; Ricardo, A.; Mahadevan, L.; Szostak, J. W. Flip-Flop-Induced Relaxation of Bending Energy: Implications for Membrane Remodeling. *Biophys. J.* **2009**, *97*, 3113–3122.

(83) Florez, C.; Raab, J. E.; Cooke, A. C.; Schertzer, J. W. Membrane Distribution of the Pseudomonas Quinolone Signal Modulates Outer Membrane Vesicle Production in Pseudomonas Aeruginosa. *MBio* **2017**, *8*.